# Tunability of the elastocaloric response in main-chain liquid crystalline elastomers

M. Lavrič[1], N. Derets[1,2], D. Črešnar[1], V. Cresta[3], V. Domenici[3], A. Rešetič[1], G. Skačej[4], M. Sluban[1], P. Umek[1], B. Zalar[1,2], Z. Kutnjak[1], and B. Rožič[1,*]

[1]*Jozef Stefan Institute, Jamova cesta 39, 1000 Ljubljana, Slovenia*
[2]*The Jozef Stefan International Postgraduate School, Jamova cesta 39, 1000 Ljubljana, Slovenia*
[3]*Dipartimento di Chimica e Chimica Industriale, Università di Pisa, via Moruzzi 13, 56124 Pisa, Italy*
[4]*Faculty of Mathematics and Physics, University of Ljubljana, Jadranska ulica 19, 1000 Ljubljana, Slovenia*

## Abstract

Materials exhibiting a large caloric effect could lead to the development of new generation of heat-management technologies that will have better energy efficiency and be potentially more environmentally friendly. The focus of caloric materials investigations has shifted recently from solid-state materials toward soft materials, such as liquid crystals and liquid crystalline elastomers. It has been shown recently that a large electrocaloric effect exceeding 7 K can be observed in smectic liquid crystals. Here, we report on a significant elastocaloric response observed by direct elastocaloric measurements in main-chain liquid crystal elastomers. It is demonstrated that the character of the nematic to paranematic/isotropic transition can be tuned from the supercritical regime towards the first-order regime, by decreasing the density of crosslinkers. In the latter case, the latent heat additionally enhances the elastocaloric response. Our results indicate that a significant elastocaloric response is present in main-chain liquid crystalline elastomers, driven by stress fields much smaller than in solid elastocaloric materials. Therefore, elastocaloric soft materials can potentially play a significant role as active cooling/heating elements in the development of new heat-management devices.

\* brigita.rozic@ijs.si

## I. INTRODUCTION

Caloric effects, such as magnetocaloric, electrocaloric or elastocaloric [1–3], are related to a reversible temperature change $\Delta T$ in a material upon applying or removing external field under adiabatic conditions. Usually, this switching process is rapid enough so that a negligible amount of heat is exchanged between the caloric material and the thermal bath. Recently, a revival of studies of the electrocaloric (ECE) and elastocaloric (eCE) effects took place due to the revelation of giant ECE response [3–14] in dielectric materials such as ferroelectrics and antiferroelectrics, as well as giant eCE response in shape memory alloys and natural rubbers [15–18]. Recent findings indicate that caloric effects have great potential for numerous heat-management applications, particularly in heating/cooling and heat waste recovery devices [1,3,6,19–21]. These new heat-management technologies are expected to provide for efficient alternative to conventional applications based on mechanical vapor compression cycle or thermoelectrics. Specifically, ECE based devices have the potential for miniaturization that can lead to a development of efficient cooling mechanisms in microelectronics. Furthermore, in contrast to the existing cooling technologies, which predominantly rely on environmentally dangerous gases or liquids, new caloric-based cooling devices will be environmentally much friendlier [19,22–24].

A typical example of a non-mechanical caloric system is a ferroelectric material, in which polarization, the order parameter of the system, is linearly coupled to the electric field, resulting in an electrocaloric response. It has been shown recently that giant electrocaloric effect can be observed in inorganic perovskite ferroelectric thin films [1,12,25], as well as in organic, P(VDF-TrFE) based ferroelectric copolymers [5,12]. Some proof-of-concept cooling devices on the basis of ECE solid state materials, were already developed, however, suffering from a rather low power density, due to the relatively large electrocaloric inactive regenerator mass [16,20,22]. The idea to replace the electrocalorically inactive fluid regenerator with the electrocalorically active dielectric fluid is driving investigations of ECE in nematic and smectic liquid crystals, where the nematic order parameter is coupled to the electric field via the dielectric anisotropy [26]. It has been shown recently that in nematic-smectic liquid crystals nCB, a significant ECE exceeding 6 K can be observed at a rather moderate electric field change of 12-80 kV/cm [26–27]. For all the cases, highest electrocaloric response was observed at the phase transition from the ordered to disordered phase accompanied by a large latent heat. Some problems with fatigue and Joule heating in ECE materials, as well as observation of giant elastocaloric response exceeding 40 K in shape memory alloys, have recently focused the attention towards mechanocaloric materials exhibiting the elastocaloric or barocaloric effect [2,15–18].

In elastocaloric materials, its stress field is coupled with the strain field. The highest elastocaloric response of 40 K was observed so far in $Ni_{50.2}Ti_{49.8}$ shape memory alloy wires at an applied stress of 0.8 GPa [2,15]. However, rather complicated experimental setups are needed in order to achieve such relatively large stress fields. This makes the miniaturization of elastocaloric cooling devices a rather complicated task. Recently, alternative elastocaloric materials, in which an order-of-magnitude smaller stress field still results in sizable entropy changes [26], were found among soft materials, specifically in liquid crystalline elastomers (LCEs) that also exhibit giant thermomechanical response [28]. Two types of LCEs are commonly produced. In side-chain LCEs (SC-LCEs), mesogens are side-attached to polymer chains of the polymer network, which is elastically stabilized by crosslinking molecules (crosslinkers) [28]. In main-chain LCEs (MC-LCEs), mesogens are themselves components of polymer chains and as such they form the polymer network [28]. In both cases, mesogens contribute the major share of the total LCE mass [28,29]. Both types of LCEs are prepared via a two stage crosslinking process, in which partially crosslinked LCEs are stretched by an external stress field. Mesogens are thus forced to form a nematic order. The internal stress field is memorized by the network during the second crosslinking stage [29]. A direct consequence of the imprinted stress memory is a large thermomechanical effect that results, by changing the temperature, in strains exceeding 400 percent in MCLCEs [28–30]. Recently, it has been demonstrated that thermomechanical response in both SC-LCEs and MC-LCEs can be tailored from on-off temperature profile to sluggish or continuous profile via changing the crosslinkers' density [28,31,32]. Lowering the density of crosslinkers yields a material with sharp, first-order type nematic transition. In contrast, increasing the density of crosslinkers above a critical value drives the nematic transition above the critical point, to the supercritical regime, with a thermomechanical response extended to a broad temperature interval [31–33]. Since, in LCEs, the external stress field is also coupled to the nematic order parameter, a significant elastocaloric response is anticipated [26]. It was shown recently that in side-chain LCEs an elastocaloric effect of $\Delta T_{eCE} = 0.4$ K can be observed already at a moderate stress field of $\sigma = 0.13$ MPa, thus demonstrating very large elastocaloric responsivity $\Delta T_{eCE}/\sigma = 4$ K/MPa [26]. These initial experimental investigations of caloric effects in LCEs [26] has been followed by a recent molecular computer simulation study [34], suggesting that the elastocaloric temperature change and responsivity observed so far in real experiments could be substantially improved.

In this work, we will address the elastocaloric effect in main-chain LCEs by direct experimental methods. One possibility to enhance the elastocaloric response is by increasing the latent heat released at the nematic-isotropic phase transition. Similarly as in the case of thermomechanical response, this could be achieved by tuning the character of the nematic phase transition, from a continuous towards a sharp first order type, by reducing the density of crosslinkers. After the experimental section, the results of Monte Carlo (MC) molecular simulations, thermomechanical response and the direct elastocaloric measurements of the elastocaloric effect in main-chain LCE will be discussed. In the final section, the summary of the above results and discussion in terms of their technological importance will be given.

## II. SAMPLE PREPARATION PROCEDURES AND EXPERIMENTAL

Elastocaloric effect was studied in three samples of main-chain liquid crystal elastomers, with respective crosslinkers' concentration of $\rho = 0.05$, 0.08 and 0.10. Here $\rho$ represents the coverage of active chain groups of the polymer backbone [31]. Specimens were prepared employing the conventional two-step Finkelmann procedure [29]. 1,1,3,3–Tetramethyldisiloxane was used as a chain extender, 2,4,6,8,10–Pentamethylcyclopentasiloxane as a crosslinker. Chemical structure of a mesogen is shown in Figure 1. Toluene and solution of dichloro(1,5–cyclooctadiene)platinum(II) in dichloromethane were used as a solvent and catalyst, respectively. The second crosslinking step was carried out in the nematic state at about 343 K. The internal stress locked-in during this step resulted in thermomechanical response of more than 100 percent. More details about sample preparation can be found in Refs. [28–30].

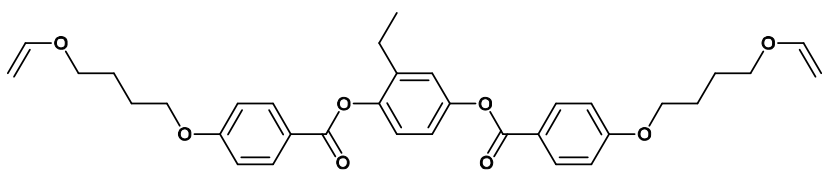

FIG. 1. Chemical formula of the mesogenic molecule used in MC-LCEs.

Mechanical testing played a significant role in the characterization of our MC-LCE specimen. Therefore, experiments were carried out with a homemade extensometer, using two regimes: (i) heating and cooling of the sample under constant force/stress and (ii) stress-strain experiment at constant temperature. The former technique allowed for the characterization of the thermomechanical response,, i.e. for the determination of the relative strain $\lambda$ *vs.* $T$ dependence,

$$\lambda(T) = \frac{l(T) - l_0}{l_0} \times 100\%.$$

Here, $l(T)$ denotes the strain at given temperature $T$ and $l_0 = l(430 \text{ K})$ denotes the reference strain well above the nematic transition where $l(T)$ stabilizes at a low value characteristic of an isotropic phase.

The eCE was evaluated by direct elastocaloric measurements. A sample of typical 15x3x0.3 mm$^3$ geometry was mounted into a homemade elastocaloric measurement setup, composed of a stress sensor, a temperature-stabilized Copper sample-chamber, and a precise step motor-driven translator, equipped with encoder providing for translation, equivalently strain, reading. The variation of the sample's temperature during elastocaloric measurements was monitored by a small-bead thermistor, attached directly to the sample.

Our molecular simulations largely follow the methodology presented in Ref. [34]. In the simulation, mesogenic units are represented by uniaxial ellipsoidal particles interacting via the soft-core Gay-Berne (GB) potential [35,36]. The model elastomeric main-chain network was grown at low density by somewhat modifying the "isotropic genesis" procedure [37] to induce an orientational bias for the polymer chains (like in the so-called Finkelmann two-step crosslinking procedure [29]) so as to reliably obtain well-oriented monodomain LCE samples in the nematic phase. The inter-particle bonds within the elastomer network were modeled by the finitely extensible nonlinear elastic (FENE) potential [38], applied both to bond stretching and bending. A limited number of tri-functional ellipsoids served as crosslinkers; in our simulations, samples with approx. 8% and 12% crosslinker content were considered. After adding the additional swelling ellipsoids (50 vol. %), the sample was isotropically compressed to close-packing. In total, there were 64000 GB ellipsoids in each sample, with periodic boundary conditions mimicking a larger bulk system. The equilibrium configurations were found by performing large-scale iso-stress Monte Carlo simulations following a modified Metropolis algorithm wherein both temperature ($T$) and external stress ($\sigma$) are fixed [39]. For more details see Ref. [34] and the references therein.

## III. RESULTS AND DISCUSSION

By now, it is very well established that the caloric effects can be greatly enhanced by increasing the latent heat released or absorbed at the first order phase transition induced by the external field during the caloric experiment [1,3,21]. Besides solid materials, such enhancement was recently demonstrated also for soft materials, i.e., smectic liquid crystals in which most of the electrocaloric effect exceeding 6 K can actually be attributed to the latent heat of the electric-field-induced isotropic-smectic A phase transition [26,27]. Similarly to solid ferroelectrics [40,41], a critical point can also be found in LCEs [28,31,32], with the useful regime of first order transition restricted to the range below the critical point in the temperature-stress field phase diagram. Above the critical point, in the supercritical regime, there is only a nematic-paranematic conversion within a broad temperature interval, without the latent heat, with strongly suppressed magnitude of the elastocaloric effect. In fact, most of the synthesized LCEs are usually found to exhibit supercritical behavior. Therefore, it is important to find the conditions in which the first order nature of the nematic phase transition is promoted. It has been shown before that the magnitude of the imprinted stress field can be controlled by the density of crosslinkers in both SC-LCEs and MC-LCEs [31–33]. Here, we substantiate this finding by additional MC simulations and measurements of the thermomechanical response.

### A. Monte Carlo simulations and the thermomechanical response

Figure 2 shows the average reduced sample side length ($\lambda$) along the imposed orientational bias direction (and, accordingly, the nematic director) as a function of reduced temperature ($T^*$) for two different crosslinking densities, $\rho = 0.08$ and 0.12. Here, $T^*$ is defined by $T^*=k_B T/\varepsilon$, where $T$ denotes the absolute temperature and $\varepsilon$ is a characteristic Gay-Berne interaction energy [36], i.e., the depth of the potential well for the parallel side-to-side molecular alignment. The less densely crosslinked sample presents a slightly lower nematic-isotropic transition temperature and a steeper temperature dependence of $\lambda$ near the nematic-isotropic phase transition in comparison with its more densely crosslinked counterpart.

Now, the elastocaloric response is dominated by the absolute value of the derivative $\left(\frac{\partial \lambda}{\partial T^*}\right)_\sigma$ [31–34], i.e., higher slopes result in a more pronounced response. In addition, a steeper slope also indicates that the transition moves toward the first order regime below the critical point in agreement with early experiments [31,32]. It should be noted that this qualitative finding of MC simulations is in a good agreement with direct measurements of the thermomechanical response in MC-LCEs at two different crosslinking densities of $\rho = 0.08$ and 0.10 shown in Figure 3. Here, Figure 3 depicts $\lambda(T)$ of the samples' first cooling run. Annealing of the samples, which removes all internal stresses, took place already during the first heating run. The rate of cooling from 430 K down to 300 K was 0.2 K/min. Both, the magnitude and the steepness of the temperature dependence of the thermomechanical response is increased with decreasing crosslinkers density, in good agreement with MC simulations. The measured and the simulated thermomechanical behavior therefore suggest that the samples with lower crosslinking densities are expected to give better elastocaloric benchmarks due to the larger absolute value of the derivative $\left(\frac{\partial \lambda}{\partial T^*}\right)_\sigma$ and possibility to drive the transition toward the first order nature with the latent heat.

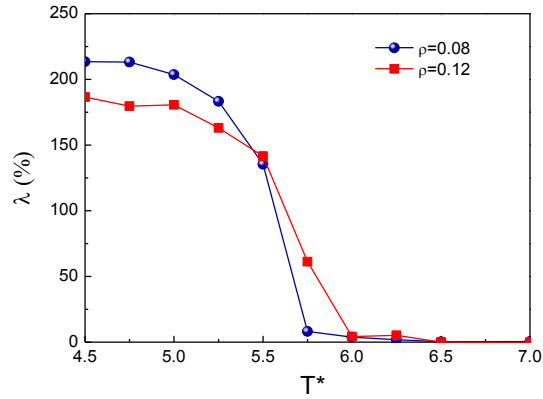

FIG. 2. Reduced sample length $\lambda$ as a function of the reduced temperature $T^*$ for two crosslinkers densities $\rho = 0.08$ and 0.12, as obtained from MC simulations.

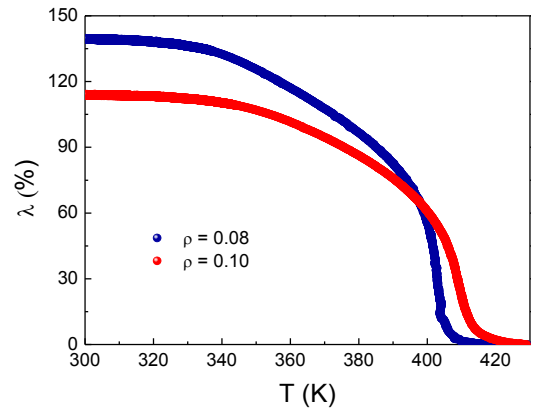

FIG. 3. $\lambda(T)$ as a function of the temperature $T$ directly measured in MC-LCEs for two crosslinkers densities $\rho = 0.08$ and 0.10.

## B. Elastocaloric response

Temperature profiles of the eCE temperature change $\Delta T_{eC}$ for MC-LCE of $\rho = 0.05$ at selected relative strains $\Delta l/l$ in the vicinity of the isotropic to nematic phase transition are shown in Figure 4. As anticipated, the maximum elastocaloric response is achieved near the phase transition taking place at about 397.5 K for an unstretched sample ($\Delta l/l = 0$). On stretching the samples ($\Delta l/l \neq 0$), the maximum is slightly shifted to higher temperatures with increasing applied stress field. In order to preserve the samples and not to exceed the breakdown field, relative strains $\Delta l/l$ were kept below 1. Nevertheless, $\Delta T_{eC} \approx 1$ K was observed at a relative strain of $\Delta l/l = 0.9$. Similar temperature profiles were also observed in MC-LCE samples of higher crosslinker compositions, i.e. for $\rho = 0.08$ and 0.10.

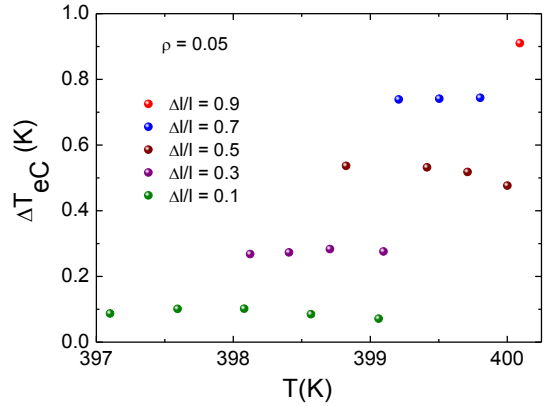

FIG. 4. The eCE temperature change $\Delta T_{eC}$ in MC-LCE of $\rho = 0.05$ as a function of temperature for several relative strains $\Delta l/l$.

Since the same value of the relative strain $\Delta l/l$ corresponds to different stress field magnitudes in MC-LCE samples of different crosslinker densities, we instead compare the stress-field dependence of the maximum elastocaloric response observed at the temperature, specifically the one corresponding to the isotropic-nematic phase transition at the given stress value. Figure 5a shows the elastocaloric temperature change $\Delta T_{eC}$ and the specific entropy change $\Delta s_{eC}$ in MC-LCE with $\rho = 0.05, 0.08,$ and 0.10 as a function of stress change $\Delta \sigma$. A slight nonlinearity in the elastocaloric response vs. stress field is observed, similarly as in the earlier observations in SC-LCE [26]. It should be noted that, on decreasing the crosslinkers density, the electrocaloric response increases significantly for a fixed stress field, in agreement with MC simulation predictions. This behavior can also be clearly observed in Figure 5b depicting elastocaloric responsivity as a function of the stress field. Even for a relatively small stress field of 0.17 MPa, $\Delta T_{eC}$ = 1.02 K was found in MC-LCE sample with $\rho$ = 0.08. Here, a higher stress field was achieved than in the even more eCE responsive sample of $\rho$ = 0.05 for which the maximal relative strain was limited to 0.9 yielding thus a somewhat lower $\Delta T_{eC}$. The observed values of $\Delta T_{eC}$ are more than one order of magnitude smaller than those observed in best solid elastocaloric materials. However, in the case of LCEs, a nearly four orders of magnitude smaller stress fields were used. Taking into account the fact that in some liquid crystalline materials large latent heat of the nematic transition [26,27] gives rise to large electrocaloric effect exceeding 6 K, and the fact that main-chain LCEs could be optimized for their strains to exceed 400 percent, it is plausible to expect elastocaloric response of more than 5 K in properly functionalized MC-LCE materials. It should as well be noted that the specific heat of MC-LCEs is typically 6 times larger than that of solid-state alloy materials, giving nearly same energy change per cooling cycle as in alloys with correspondingly larger $\Delta T_{eC}$.

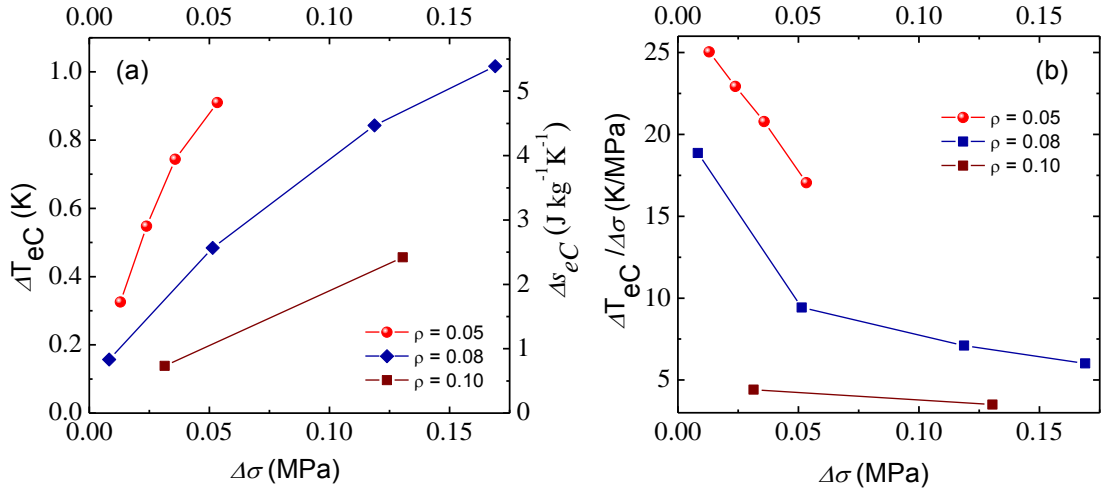

FIG. 5. (a) The elastocaloric temperature change $\Delta T_{eC}$ and the elastocaloric specific entropy change $\Delta s_{eC}$ in MC-LCE of $\rho$ = 0.05, 0.08 and 0.10 as a function of stress-field change $\Delta\sigma$. (b) Elastocaloric responsivity $\Delta T_{eC}/\Delta\sigma$ in MC-LCE of $\rho$ = 0.05, 0.08 and 0.10 as a function of stress-field change $\Delta\sigma$.

Moreover, the maximum value of the elastocaloric responsivity $\Delta T_{eC}/\Delta\sigma$ = 25 K/MPa, observed in MC-LCE of $\rho$ = 0.05, is more than two orders of magnitude larger than the average elastocaloric responsivity of ≈0.04 K/MPa found in best shape memory alloys [2].

This makes MC-LCEs with low crosslinking densities and large latent heat highly promising candidates for the soft elastocaloric materials with large elastocaloric response.

**IV. CONCLUSION**

In conclusion, we have investigated, by direct experiments, the elastocaloric effect in main-chain liquid crystal elastomers of different crosslinking compositions. By comparing numerical predictions of Monte-Carlo simulations and experimental results, we showed that the density of crosslinkers can tailor the magnitude of the elastocaloric response, by shifting the character of the nematic transition from the smeared supercritical towards the sharp first order type. The existence of a sizable elastocaloric effect exceeding 1 K is demonstrated in main-chain liquid crystal elastomers, with a more than two orders of magnitude larger elastocaloric responsivity $\Delta T_{eC}/\Delta\sigma$ = 25 K/MPa than that found in shape memory alloys champions. It is argued that in MC-LCEs with large latent heat and large strains exceeding 400 percent, it should be possible to achieve elastocaloric responses of more than 5 K, which, together with the large specific heat of MC-LCEs, could provide nearly same energy change per cooling cycle as other caloric materials.

The data that supports the findings of this study are available within this article.

**ACKNOWLEDGEMENTS**

This work was supported by the European Union project 778072 — ENGIMA — H2020-MSCA-RISE-2017, European Regional Development Fund; NAMASTE Centre of Excellence, Space-SI Centre of Excellence, Slovenian Research Agency grants J1-9147, and programs P1-0125 and P1-0099.